\documentstyle[preprint,eqsecnum,aps,epsfig]{revtex}

\tightenlines

\newcommand{\be}{\begin{equation}}    % for lazy typers
\newcommand{\ee}{\end{equation}}
\newcommand{\beq}{\begin{eqnarray}}
\newcommand{\eeq}{\end{eqnarray}}
\newcommand{\beqn}{\begin{eqnarray*}}
\newcommand{\eeqn}{\end{eqnarray*}}

\def\msun{M_\odot}

\def \tablerule{\noalign {\vskip3truept\hrule\vskip3truept}}    
\def \up {0pt}

\begin{document}

%%%%%%%%%%%%%%%%%%%%%%%%%%%%%%%%%%%%%%%%%%%%%%%%%%%%%%%%%%%%%%%%%%%%%%
% uncomment the following two lines and one below for two columns!
%%%%%%%%%%%%%%%%%%%%%%%%%%%%%%%%%%%%%%%%%%%%%%%%%%%%%%%%%%%%%%%%%%%%%%

%@twocolumnfalse\endcsname

\title{
Are Post-Newtonian templates faithful and effectual in detecting gravitational signals
from neutron star binaries?}
%emitted in the inspiralling of neutron stars  binary systems?}

\author{E. Berti$^{1}$, J.A. Pons$^{2}$, G. Miniutti$^{2}$,
L. Gualtieri$^2$ and V. Ferrari$^{2}$}

\address{
$^1$ Department of Physics, Aristotle University of Thessaloniki,
Thessaloniki 54006, Greece}

\address{$^2$ Dipartimento di Fisica ``G.Marconi",
 Universit\` a di Roma ``La Sapienza"\\
and Sezione INFN  ROMA1, piazzale Aldo  Moro
2, I-00185 Roma, Italy}

\maketitle

\begin{abstract}
We compute  the overlap function between Post-Newtonian (PN) templates and 
gravitational signals emitted by   binary systems composed of one neutron 
star and one point mass, obtained by a perturbative approach. 
The calculations are performed
for different stellar models and for different detectors,
to estimate how effectual and faithful the PN templates are,
and to establish whether effects related to the internal structure of neutron stars
may possibly be extracted  by the matched filtering technique.
\end{abstract}


\vskip2pc

%%%%%%%%%%%%%%%%%%%%%%%%%%%%%%%%%%%%%%%%%%%%%%%%%%%%%%%%%%%%%%%%%%%%%%%
% Introduction
%%%%%%%%%%%%%%%%%%%%%%%%%%%%%%%%%%%%%%%%%%%%%%%%%%%%%%%%%%%%%%%%%%%%%%%

\section{Introduction}
The detection of gravitational waves  emitted  during
the late inspiral and merger phases of coalescing compact binaries 
is one of  the main targets of the ground based interferometric  detectors 
that are currently in the final stage of construction or in
the commissioning phase (LIGO, VIRGO, GEO600, TAMA).
To detect these signals, a detailed knowledge of the emitted waveforms 
is fundamental; indeed, the performances of the matched filter which 
will be used to extract the signal from
the detectors noise, largely depend on the capability of the 
theoretical templates to reproduce the true waveforms. 

Although the population of  neutron star-neutron star (NS-NS) binaries
is expected to double that of black hole-black hole (BH-BH) binaries,
and to be several times larger than that of (NS-BH) \cite{binaries},
BH-BH binaries having a total mass  of
$\sim(20-40)$ M$_\odot$ will likely    be detected first by
the initial ground-based interferometers, because, due to their larger mass,
the signal is more intense in the frequency region where the 
detectors are more sensitive \cite{grishchuk}.
In the future, however, as the detectors  sensitivity in the
high frequency region improves,
NS-NS coalescence should become  detectable as well. 
According to recent investigations \cite{kojima,holai,tutti1,tutti2},
the signal emitted during the latest phases preceeding coalescence differs
from that emitted by two black holes essentially in one respect: the modes of
oscillation of the stars  could be ``marginally excited".
A mode is resonantly excited if the system moves on 
an orbit such that the Keplerian orbital frequency, $\omega_k,$ is in a 
definite ratio with the mode frequency $\omega_i,$ i.e. if 
$\ell\omega_k=\omega_i,$ where $\ell$ is the harmonic parameter.
In general the frequency of the fundamental mode is too high to be excited directly, 
because the stars merge before reaching the corresponding  orbit \cite{holai}; 
however, the width of the resonance of the fundamental mode (especially for $\ell=2$)
is large enough to allow the mode to be  marginally excited before  the resonant 
frequency is reached.
As a consequence, more energy is  emitted with respect to that due
to the orbital motion,  the process of inspiralling is accelerated, and 
this changes the phase of the emitted signals during the last orbits before merging. This
effect is stronger for stiffer equations of state (EOSs), or for low mass NSs,
for which the frequency of
the  fundamental mode is lower, and the  width of the resonance is larger
\cite{tutti2}.
Thus, an accurate detection of the signal emitted in a NS-NS binary coalescence
besides probing  the theory of gravity, as BH-BH signals would do, 
would also give an insight into the equation of state of matter
at high density regimes unreachable in a terrestrial 
laboratory.  

The aim of this paper is to investigate whether the templates that 
are being constructed to extract the signal emitted by  inspiralling 
binaries  from the detectors noise
are well suited to detect a NS-NS coalescence.

%%%%%%%%%%%%%%%%%%%%%%%%%%%%%%%%%%%%%%%%%%%%%%%%%%%%%%%%%%%%%%%%%%%%%%%%%%
\section{ Perturbative approach versus post-newtonian expansions. }
%%%%%%%%%%%%%%%%%%%%%%%%%%%%%%%%%%%%%%%%%%%%%%%%%%%%%%%%%%%%%%%%%%%%%%%%%%
The basic functions needed to construct the gravitational wave
(GW) templates are the  orbital energy of the system, $E(v),$
and  the gravitational luminosity,  
$\dot E_{GW}(v)$, where $v$ is the orbital velocity.
We evaluate $\dot E_{GW}(v)$ by using a perturbative approach, assuming that 
one of the two bodies is a neutron star
(hereafter we shall refer only to non rotating stars and black holes),
whose equilibrium structure is described by a solution of 
the relativistic equations of  hydrostatic equilibrium;
the second body  is a test-particle in circular orbit,
which  induces a perturbation on the gravitational field and on the thermodynamical 
structure of the extended companion.  
By this approach, we can account for the relativistic tidal effects of the close
interaction on one of the two stars, and for the effects that the
internal structure  has on the gravitational emission.
Since the particle moves on a geodesic,
its orbital energy normalized to the mass of the star, $E(v),$ is known
\be
\label{e(v)}
E(v)= \eta~ \left(1-2v^2 \right)\left(1-3v^2 \right)^{-1/2};
\ee
in this formula
$\eta=m_1/m_2,$  $m_2$  is the mass of the central body,
$m_1 \ll m_2$ is the particle mass,
$v=(\pi m_2 \nu_{GW})^{1/3}$ is the orbital velocity, 
and $\nu_{GW}$ is the frequency of the emitted radiation.
It is useful to define the normalized GW-luminosity
that  will be used in the following
\be
\label{pdiv}
P(v)\equiv \dot E_{GW}(v)/\dot E_N(v),
\ee
where $\dot E_N(v)=32\eta^2 v^{10}/5$
is the Newtonian quadrupole luminosity.
In \cite{tutti1,tutti2} we solved the  relativistic equations of stellar perturbations
and computed $\dot E_{GW}(v)$ during the final 
stages of inspiralling  of this idealized binary system.
In \cite{tutti2} we considered 5 models of polytropic stars, 
labelled from A to E, with parameters
chosen to cover most of the range of structural properties obtained with realistic EOSs. 
These parameters are given in Table \ref{table1}.
In the following, we shall refer  to the GW signals computed for the different models as
the {\it true} signals.

In the standard PN approach,  $E(v)$ and $\dot E_{GW}(v)$   are found
by assuming that both compact objects are pointlike, and by expanding 
the general relativistic equations of motion and the wave generation 
formulas in powers of $v/c$.  In the test-particle limit $E(v)$ is 
known exactly (Eq. \ref{e(v)}); $\dot E_{GW}(v)$ has been derived by Taylor 
expanding up to $(v/c)^{11}$
the solution of the Bardeen-Press-Teukolsky (BPT) equation 
\cite{bardeenpress,teukolski}, describing the perturbations
of a Schwarzschild black hole driven by an orbiting test-particle
\cite{CutlerFinnPoissonSussman,poisson47,tanakatagoshisasaki}.

In the case of comparable masses, the  equations of binary motion have 
been computed at the 3PN order, and the energy flux has been evaluated 
without ambiguities up to 2.5PN order with respect to the quadrupole formalism
(PN calculations to order $(v/c)^7$, developed by currently 
used techniques, leave undetermined a parameter entering at order $(v/c)^6$ \cite{blanchet}). 
Important progresses have recently been made
by the introduction of re-summation techniques 
(see \cite{DIS1} and references therein), 
which improve the convergence of the PN series 
(the Taylor expansion is indeed rather poorly convergent), and 
new filters in the frequency domain have been proposed,
which combine the performance of Pad\'e-approximants with the simplicity of 
the stationary phase approximation. 
Summarizing, the PN approach allows to compute the phase evolution 
of the GW signal emitted by a binary system 
to order $(v/c)^{11}$ in the test particle limit, and to order  $(v/c)^5$ 
in the case of comparable masses.

Since tidal effects do not affect the evolution of a BH-BH binary system, 
even when the black holes have comparable masses \cite{Alvi},
PN expansions are particularly well suited to evaluate the waveform in the 
case of BH-BH inspiralling.  However, when at least one of the 
coalescing bodies is a neutron star, the effects that
its internal structure may have on the gravitational emission, such as 
the modes excitation, have not been investigated in great detail
until recently. Thus, the following question arises. Suppose that, by using the 
PN approach described above we construct a GW template
and suppose that the gravitational event  involves at least one neutron 
star: what would we miss in using the PN template in the data analysis? 
Would the signal to noise ratio deteriorate because
we are using an unappropriate template, or would it be good enough 
to detect the signal anyway?
What would be the error in estimating  the binary parameters?
Or, using the terminology introduced in \cite{DIS1}, how {\it effectual} 
and {\it faithful} is to use BH templates when the signal comes from 
NS binaries? 

It should be stressed that, to date, there is no fully non linear, 
dynamical simulation of the inspiralling and coalescence of stars or 
black holes with comparable masses, which provides an exact waveform.
For this reason  the calibration and convergence tests of  PN expansions 
for  comparable masses have been done as follows \cite{DIS1}:
the functions $P(v)$ is computed 1) by integrating the BPT equation for a perturbed black hole
excited by an orbiting test particle  and 2) by computing the PN expansion 
of the function to the desired order, and ignoring the dependence of the 
expansion coefficients on the symmetric mass ratio
$\eta=(m_1\cdot m_2)/(m_1+m_2)^2$ (note that if $m_1 \ll m_2,$
$\eta \rightarrow m_1/m_2$).
In this way, the $\eta$-dependence is 
kept only at the leading order. The two signals are then compared.
The effectualness of the PN templates is computed by a 
{\it naive approximation}, by using the signal
obtained by integrating the BPT equation as a true signal, and the PN approximants as 
templates, allowing $\eta$ to assume a finite value (see for instance Table V in 
\cite{DIS1}).

In a similar way, to answer the questions above  we shall use the PN templates
obtained  as in  2) and replace 1) by the integration of the equations of stellar
perturbations excited by the orbiting test particle.
We hope that the lessons we learn from the results of this comparison
will be useful when more accurate models for
the signals emitted by real binary systems will be available.
In addition, since no generalization of the perturbative calculations
to the equal mass case is presently available,
establishing whether or not the structural effects are expected to be relevant 
from the point of view of detection would provide a motivation for 
looking for a suitable generalization of the perturbative approach.

%%%%%%%%%%%%%%%%%%%%%%%%%%%%%%%%%%%%%%%%%%%%%%%%%%%%%%%%%%%%%%%%%%%%%%%
\section{Effects of neutron star structure. }
%%%%%%%%%%%%%%%%%%%%%%%%%%%%%%%%%%%%%%%%%%%%%%%%%%%%%%%%%%%%%%%%%%%%%%%
As an example of the effects that the internal structure of the neutron star
has on the emitted signal, in Fig. \ref{fig1}a) we show the normalized GW-luminosity
(\ref{pdiv}), $P(v)$,  as a function of the orbital velocity, computed for a BH   and
for a neutron star  perturbed by a test-particle in circular orbit.
The curve for the star refers to a NS of mass  $1.4~\msun$ and radius  $R=15~$km,
labeled as model B  (see Table \ref{table1}).
The gravitational luminosity $\dot E_{GW}(v)$ is
obtained by solving the equations of stellar perturbations as described in
\cite{tutti2}, and the BPT equation for a Schwarzschild black hole excited by the same
source.
For comparison, in the same figure we plot the function $P(v)$ as computed by the 
PN formalism at order 2.5 (both Pad\'e and Taylor), and at order 5.5 (only Taylor).
For a definition of Taylor and Pad\'e approximants we refer to  \cite{DIS1}.

When the central object is a neutron star, the resonant excitation of
the quasi-normal modes (the peaks in Fig. \ref{fig1}a)
clearly changes the shape of the function $P(v)$ with respect to the BH case, 
and the effect starts to be seen at orbital velocities larger than 0.18 
\cite{tutti2}.  We see that P-approximants   rapidly converge  
to the BH solution in the test-particle limit.
As discussed in \cite{DIS1},
T-approximants of order 2.5 are especially bad-behaved, but
P-approximants seem to reproduce very well the function $P(v)$
for a BH, even at the  2.5 PN order. 
Templates built in this way are fast and convenient to be used in the 
laborious matched-filtering process.
However, for sufficiently high order, both T- and P-approximants converge to the
BH solution by construction, since $P(v)$ is computed 
as a series expansion about the exact numerical solution of the BPT equation.

In order to appreciate  the differences between the emission of different stellar
models, in Fig. \ref{fig1}b) we show  the function $P(v)$ computed for the 
models given in Table \ref{table1},
restricted to the region $v < 0.28,$ which corresponds to a distance between the two stars
larger than 2.5-3 stellar radii depending on the model. The normalized energy fluxes emitted by
different stellar models have a different slope, and are always larger than the
flux emitted by the black hole, which is also shown for comparison.
The curve for  model E  is practically
indistinguishable from the black hole  curve and that for  model D 
is also very close to  the black hole result. 
The  steepest raise of the curves
of  models  A,B,C is   a marginal effect of the excitation of the quasi normal modes.

It is clear from  Fig. \ref{fig1}b) that the differences are small,
and whether or not  they could play a role in the detection 
depends on the characteristics of the detector. This issue is discussed
in a more quantitative way in the next sections.

%%%%%%%%%%%%%%%%%%%%%%%%%%%%%%%%%%%%%%%%%%%%%%%%%%%%%%%%%%%%%%%%%%%%%
\section{Faithfulness and effectualness}
%%%%%%%%%%%%%%%%%%%%%%%%%%%%%%%%%%%%%%%%%%%%%%%%%%%%%%%%%%%%%%%%%%%%%
We shall now evaluate how  effectual and faithful the PN templates are in detecting a
signal emitted by an inspiralling neutron star binary system. 
We shall assume that the functions $\dot E_{GW}(v)$ and
$E(v)$ are known both for the PN template and for the {\it true signal} obtained by
the perturbative approach.
The two waveforms $h^A(t)$ (the PN {\it template}) 
and $h^X(t)$ (the {\it true} NS signal) will be computed in the 
so-called restricted PN approximation: the amplitude of the waveform is 
taken at the lowest PN order, while the orbital phase evolution
of the binary during the quasi-stationary inspiralling
is evaluated by numerically integrating the equations
\begin{eqnarray}
\frac{dv}{dt} &=& - \frac{1}{M_{tot}} \frac{\dot E_{GW}(v)}{dE(v)/dv} \\
\frac{d\phi}{dt} &=& \frac{2}{M_{tot}} v^3,
\end{eqnarray}
where $M_{tot}=m_1+m_2$.
Explicitly, the waveform from a source at a distance $r$ from Earth is:
\be
r h(t) = 4 C \eta M_{tot} v^2(t) \cos{\phi(t)},
\ee
where $C$ is a constant which depends on the relative orientation of the source and the
detector \cite{sathyamothertemplates}.
Let us define the Wiener inner product between the waveform $h^A$ and
$h^X$, shifted by a time lag $\tau$, as
\be\label{innprod}
\langle h^A,h^X \rangle=\int_{-\infty}^{\infty}
{d\nu e^{2\pi i \nu \tau}\over S_N(\nu)}\tilde{h}^A(\nu)~\tilde{h}^{X*}(\nu),
\ee
where $S_N(\nu)$ is the one-sided power spectral density of the detector's noise, 
and $\tilde{h}^A,\tilde{h}^{X}$ are
the Fourier transforms of $h^A(t),h^X(t)$. We compute $\tilde{h}$
using standard FFT routines because we prefer to avoid further 
uncertainties introduced when $\tilde{h}^{A,X}$ are obtained in 
the stationary phase approximation.
The noise curves of LIGO I and VIRGO I which we are going to use 
are taken from \cite{DIS3}, and those  of EURO and EURO-Xylophone, 
the third-generation interferometric
detector recently proposed, from \cite{euro}.
The explicit expressions of $S_N(\nu)$ are given in Table \ref{table2}. 

From the Wiener scalar product and the associated norm we can build the
so-called ambiguity function
\be
{\cal A}=
{\rm max}_{\tau,\Phi^A,\Phi^X}~{\langle h^A,h^X\rangle \over ||h^A|| ~||h^X||}
={\rm max}_{\tau,\Phi^A,\Phi^X}~{\langle \hat h^A,\hat h^X\rangle}.
\ee
where we denote by a hat the normalized waveforms: $\hat h^A=h^A/||h^A||$, and
$\hat h^X=h^X/||h^X||$.

For simplicity, we shall assume that the approximate and exact waveforms
$h^A$ and $h^X$ depend only on their initial phases and on the masses
of the binary members (a more realistic parameterization would require
the inclusion of spins, angular dependences, etc.):
$$h^A=h^A(t,\Phi_c^A,m_1,m_2), \qquad h^X=h^X(t,\Phi_c^X,m_1,m_2).$$

The {\it faithfulness} is defined as
\be
F_{AX}={\rm max}_{\tau, \Phi_c^A, \Phi_c^X}
~\langle \hat h^A(\tau,\Phi_c^A,m_1,m_2), \hat h^X(0,\Phi_c^X,m_1,m_2)\rangle,
\ee
i.e., as the maximum of the ambiguity function over the initial phases 
and the time lag, when the source and template parameters (in our case the masses)
are matched.
To maximize over the difference in times of arrival $\tau$ and over
the phases we follow the approach described in the Appendix A
of Ref. \cite{DIS1}.

In practice, in a detection the source  parameters are unknown.
The maximum of the ambiguity function with respect to phases and 
times of arrival will be smaller than one, and will occur, in general, when
the parameters of the source and template are not equal. 
In this case it is useful to compute
the {\it effectualness}, i.e. to maximize the ambiguity function
over all the parameters of the template
\be
E_{AX}={\rm max}_{\tau, \Phi_c^A, \Phi_c^X, m_1^A, m_2^A}~
\langle \hat h^A(\tau,\Phi_c^A,m_1^A,m_2^A), 
\hat h^X(0,\Phi_c^X,m_1^X,m_2^X)\rangle.
\ee

Thus, the computation of the {\it effectualness} differs from that of the 
{\it faithfulness} essentially in one respect: we maximize not only over the 
difference in times of arrival $\tau$ and over the phases of the waves, 
but also over the masses
of the template waveform $m_1^A$, $m_2^A$, or, equivalently, over the 
symmetric mass ratio $\eta^A=m_1^Am_2^A/(M_{tot}^A)^2$ and over the total 
``chirp mass" ${\cal M^A}=(\eta^A)^{3/5} M_{tot}^A$.

A reasonable criterion for a template to be effectual is that the ambiguity 
function thus computed should be larger than 0.965, which ensures that no 
more than 10 \% of the events are lost (Number of lost events= $1-E_{AX}^3$).

%%%%%%%%%%%%%%%%%%%%%%%%%%%%%%%%%%%%%%%%%%%%%%%%%%%%%%%%%%%%%%%%%%%%%%%
\section{Results and Concluding Remarks}
%%%%%%%%%%%%%%%%%%%%%%%%%%%%%%%%%%%%%%%%%%%%%%%%%%%%%%%%%%%%%%%%%%%%%%%

\begin{figure}
\begin{center}
\leavevmode
\epsfxsize=3.5in \epsfbox{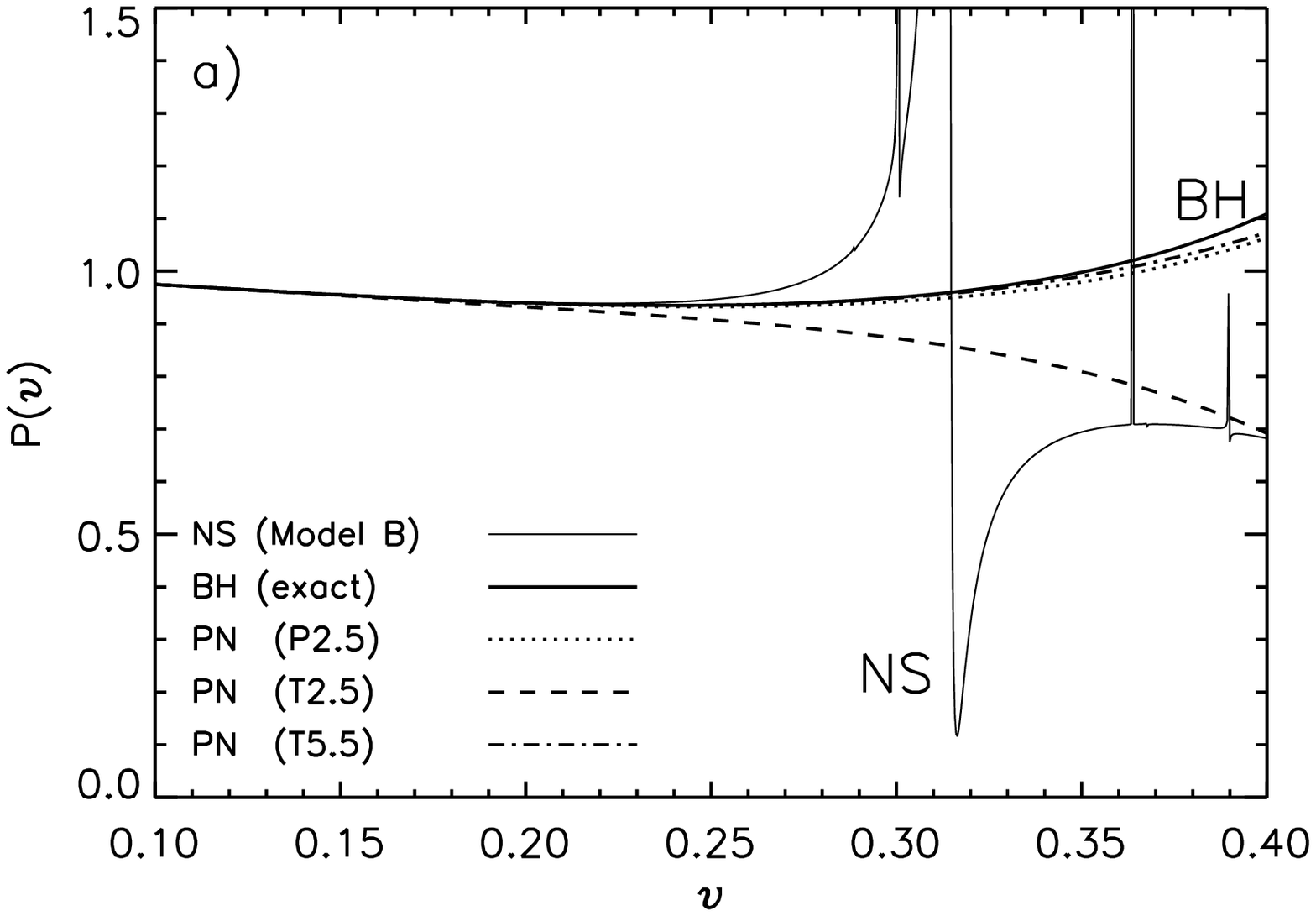}~~~
\epsfxsize=3.5in \epsfbox{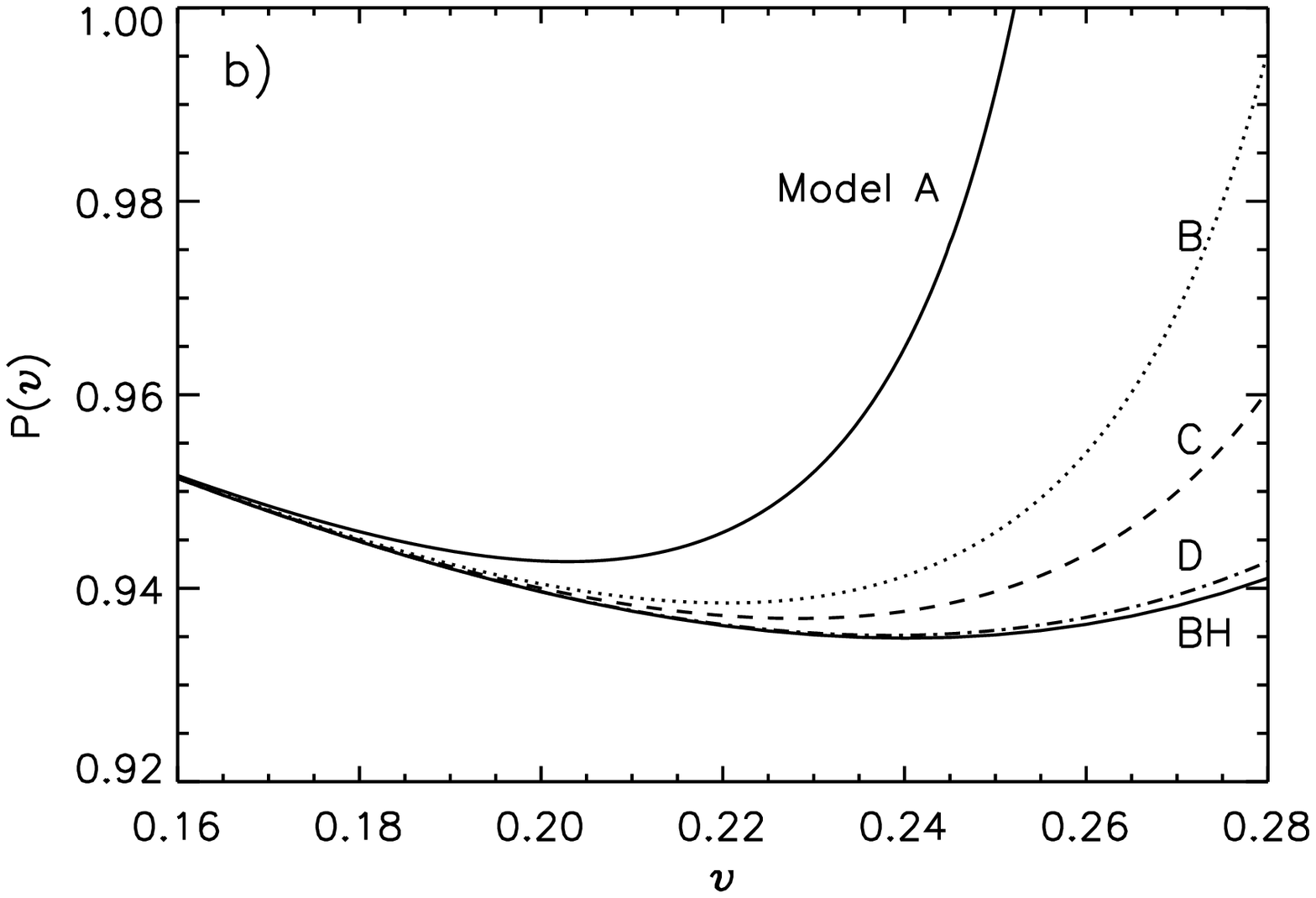}
\end{center}
\caption{
The normalized GW-luminosity,   $P(v),$ is plotted versus the orbital
velocity. On the left, we show  the curve for the stellar model B
and for a black hole.
For comparison, we also show the 2.5 PN Pad\'e and Taylor  approximants,
and the 5.5 PN Taylor approximant.
The peaks correspond to the excitation of the  $f-$mode of the neutron star
for different $l$s, and the wider resonance corresponds to $l=2$.
On the right,  we plot   $P(v)$ for all stellar models  given in Table \ref{table1}
and for a black hole,  for a smaller  orbital velocity  range.
}
\label{fig1}
\end{figure}

\begin{figure}
\begin{center}
\leavevmode
\epsfxsize=4in \epsfbox{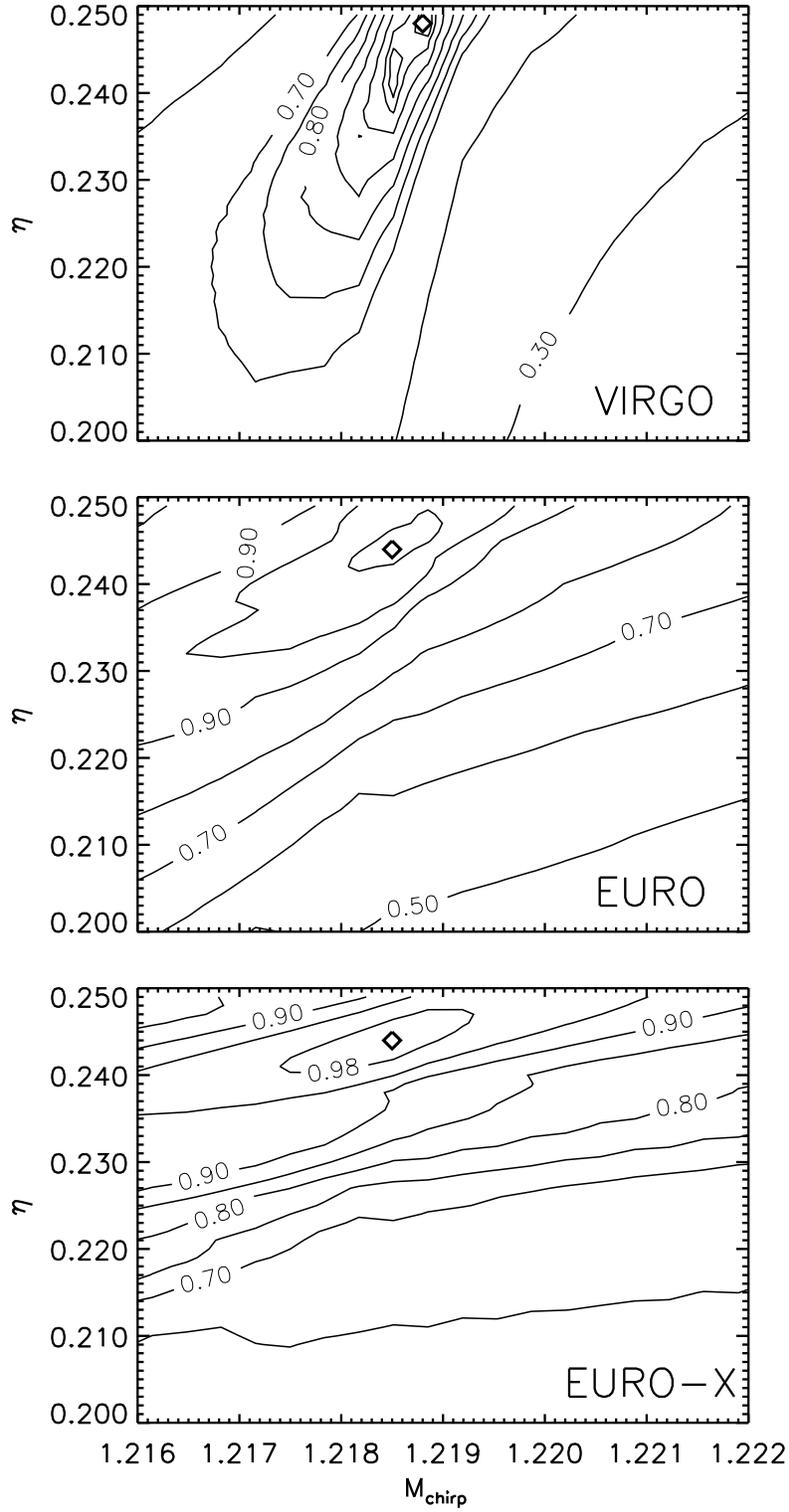}
\end{center}
\caption{Constant ({\it minimax}) overlap levels are plotted as a function of the chirp 
mass ${\cal M}$ and of the symmetric mass ratio $\eta$.
The {\it true} signal is emitted by
the stellar model D with an orbiting test particle of $1.4 M_\odot$  (the true values of 
$\eta$ and ${\cal M}$ are $\eta=0.25,$ and ${\cal M}=1.2188 M_\odot$),
the PN template is Pad\'e-6 and the noise curves are those of VIRGO, EURO and
EURO-Xylophone (EURO-X). 
The diamond indicates the maximum of the ambiguity function.
}
\label{FIG2}
\end{figure}

In Fig. \ref{FIG2} we show the contour levels of constant  ambiguity function,
as a function of ${\cal M}$ and $\eta$. 
We compute the {\it minimax} overlap, which is physically more
relevant for detection than the {\it best} overlap, as explained in
\cite{DIS1}.  
The {\it true} waveform is that emitted by
the stellar model D excited by  an orbiting point particle of $1.4 M_\odot$,
while the PN template  is Pad\'e-6. The noise curves are, respectively,  those of VIRGO I,
EURO and EURO-Xylophone, indicated as EURO-X. 
We do not plot the curves for LIGO I, because they are very similar to those of VIRGO I.
The true parameters of the  binary system are $\eta=0.25$ and 
${\cal M}=1.2188 M_\odot$.
Model D is one with a soft EOS, and it is very compact. 
The frequency of the fundamental mode is too high to be excited at a significant 
level, and indeed in Fig. \ref{fig1}b)
we see that its GW-luminosity is very close  to that of a black hole.
Thus, we expect that in this case both the chirp mass and $\eta$ will 
be accurately determined by the PN templates, that
are constructed for black hole coalescence. This is confirmed in
Fig. \ref{FIG2} for all the considered  detectors.
From the three panels of the figure we see that the most important parameter
in determining the overlap function is the chirp mass, which can
be inferred with an error smaller than one part in a thousand
(notice the scale on the $x$-axis).  This fact was already noted in refs.
\cite{cutlerflanagan,grishchuk}.  The dependence
on the symmetric mass ratio is found to be somewhat weaker, and the relative error in its 
estimation is about 2-3 \%.

\begin{figure}
\begin{center}
\leavevmode
\epsfxsize=4in \epsfbox{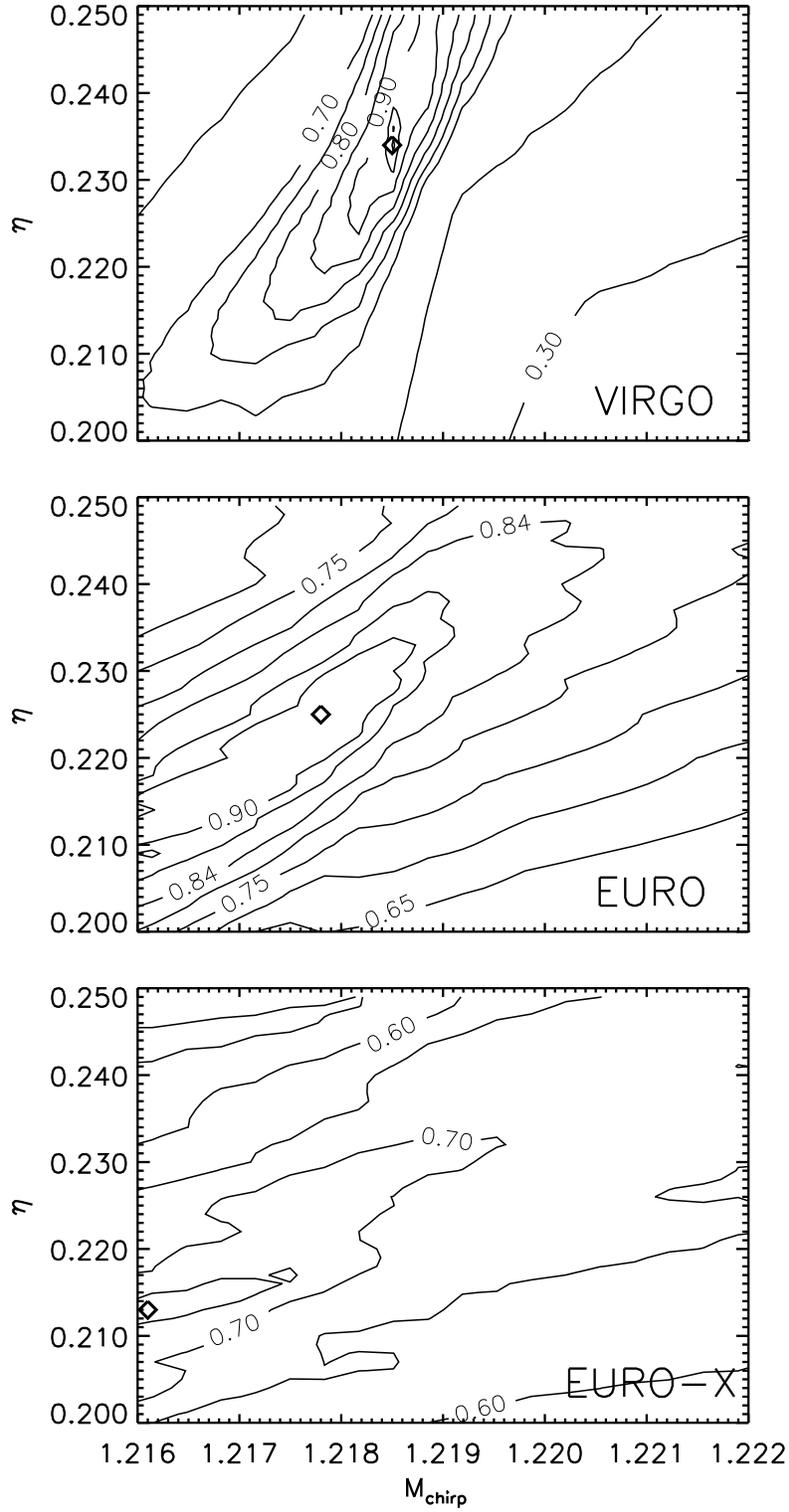}
\end{center}
\caption{Constant ({\it minimax}) overlap levels  are plotted as in Fig. \ref{FIG2},
with the {\it true} signal emitted by the stellar model B + orbiting test particle
(true values: $\eta=0.25,$ and ${\cal M}=1.2188 M_\odot$), and the noise curve of VIRGO,
EURO and EURO-Xylophone. 
}
\label{fig3}
\end{figure}

It is interesting to plot a similar figure  for the 
stellar model B.
This model has a stiff EOS, and the marginal excitation of the fundamental
mode before merging is visible in Fig. \ref{fig1}b).
From Fig. \ref{fig3} we see that if the detector is VIRGO,
the chirp mass and the mass ratio where  ${\cal A}$ has a maximum
are ${\cal M}= 1.2185 M_\odot$ and $\eta=0.234;$
if the detector is EURO, ${\cal M}=1.2178  M_\odot$ and $\eta=0.225,$
and if the detector is EURO-Xylophone  ${\cal M}= 1.2161 M_\odot$ and $\eta=0.213$.
Thus, whereas the chirp mass would still be determined to a very  good accuracy
the determination of $\eta$ would be less accurate if the detectors are EURO-type,
i.e. very sensitive at high frequency,
and  if the templates remain tuned to the black hole signal.
This can be understood also by looking at 
Table \ref{table3}, where we give  the values of the  effectualness  for
LIGO I, VIRGO I, EURO and EURO-X. The {\it true} signal is that emitted by the
five models of NSs given in Table \ref{table1}; as a template we
use the Pad\'e-6 approximant (column 1 for each detector), and the
signal obtained by integrating the BPT equation for a Schwarzschild
black hole perturbed by an orbiting particle (column 2), because the
Pad\'e approximant is nothing but an approximation of this signal;
thus we wanted to check what is the change if we use as a template the
exact signal emitted by a black hole.
We see that the performances of $P$-approximants,
as well as those of BH approximants, degrade if the
detector is very sensitive in the high frequency region.
In this case, the use of templates which account for effects
of stellar structure would be needed.
In Table \ref{table4} we show the analogous results for the faithfulness.

The required effectualness threshold  of 0.965 is always 
achieved in LIGO I for all of the stellar models;
for VIRGO I  it is a little lower
because the detector is more sensitive at high frequency.
Thus, if the coalescing binary system is composed of two neutron stars,
both LIGO and VIRGO would be able to detect it by using the standard PN templates,
and to determine the masses with a sufficient accuracy, provided
the event occurs  close enough to be visible by these instruments.

If the noise curve is that of EURO or EURO-Xylophone,
the effectualness is lower and a relevant fraction of events would 
be missed using the standard PN templates; 
for instance EURO would miss $\sim$ 36\% of the events if
the stars have low mass as in the stellar model A,
and  $\sim$ 18\% if the EOS is that of model B. For EURO-Xylophone it would be worse:
$\sim$ 78\%  events missed for model A,  $\sim$ 57\% for model B and $\sim$ 33\% for model
C.
We would like to emphasize that this difference between neutron star 
models is what makes the situation more interesting.
By constructing  filters which  include resonant effects,
we would both increase the chances to
detect NS-NS events and  be able to estimate the oscillation 
frequencies of NSs.  Knowing the mass of the stars, these could be used to infer 
their radius, as suggested by recent investigations \cite{asterosysm,vallisneri},
and set constraints on the EOS of nuclear matter
in the supranuclear density regime \cite{lattimer}.
In addition, it should be stressed that
the imprint that the internal structure of the stars leaves 
on the GW signal may be enhanced by rotation,
the effect of which is to  lower some of the mode frequencies; 
this would shift  the  effect of the mode excitation toward lower frequencies 
and amplify the signal in the region where EURO-type detectors  would be more sensitive.  
\acknowledgments
We would like to thank Dr.  B.S. Sathyaprakash for useful discussions
on the use of PN templates and for providing us the correct noise curve of EURO.

This work has been supported by the EU Programme `Improving the Human
Research Potential and the Socio-Economic Knowledge Base' (Research
Training Network Contract HPRN-CT-2000-00137).
JAP is supported by the Marie Curie Fellowship No. HPMF-CT-2001-01217.

\newpage

%%%%%%%%%%%%%%%%%%%%%%%%%%%%%%%%%%%%%%%%%%%%%%%%%%%%%%%%%%%%%%%%%%%%%%%%%%
%%%%%%%%%%%%%%%%%%%%%%%%%%%%%%%%%%%%%%%%%%%%%%%%%%%%%%%%%%%%%%%%%%%%%%%%%%%
%%%%%%%%%%%%%%%%%%%%%%%%%%%%%%%%%%%%%%%%%%%%%%%%%%%%%%%%%%%%%%%%%%%%%%%%%%%
%%%%%%%%%%%%%%%%%%%%%%%%%   TABLES   %%%%%%%%%%%%%%%%%%%%%%%%%%%%%%%%%%%%%
%%%%%%%%%%%%%%%%%%%%%%%%%%%%%%%%%%%%%%%%%%%%%%%%%%%%%%%%%%%%%%%%%%%%%%%%%%%
%%%%%%%%%%%%%%%%%%%%%%%%%%%%%%%%%%%%%%%%%%%%%%%%%%%%%%%%%%%%%%%%%%%%%%%%%%%%%%%%
\begin{table}
\centering
\caption{
Parameters of the polytropic stars we consider in our analysis:
the polytropic index $n$, the central density, the ratio
$\alpha_0=\epsilon_0/p_0$ of central energy
density to central pressure, the mass,  the radius  and the ratio $M/R$
($\alpha_0$ and $M/R$ are in geometric units).  The central energy density
is chosen in such a way that the stellar mass is equal to
$1.4 M_\odot$, except for model A, the mass of which is about one solar mass.
}
\vskip 12pt
\begin{tabular}{@{}clllllr@{}}
\hline
Model number &$n$ &$\rho_c$ (g/cm$^3$) &$\alpha_0$  &$M$ ($M_\odot$) &$R$ (km)&
$M/R$\\
\hline
A   &1.5     & $1.00\times 10^{15}$    &13.552     &$0.945$     &14.07 & 0.099\\
B   &1       & $6.584\times 10^{14}$   &9.669      &$1.4$       &15.00 & 0.138\\
C   &1.5     & $1.260\times 10^{15}$   &8.205      &$1.4$       &15.00 & 0.138\\
D   &1       & $2.455\times 10^{15}$   &4.490      &$1.4$       &9.80  & 0.211\\
E   &1.5     & $8.156\times 10^{15}$   &2.146      &$1.4$       &9.00  & 0.230\\
\hline
\end{tabular}
\label{table1}
\end{table}
%%%%%%%%%%%%%%%%%%%%%%%%%%%%%%%%%%%%%%%%%%%%%%%%%%%%%%%%%%%%%%%%%%%%%%%%%%%%%%%%%%%

%%%%%%%%%%%%%%%%%%%%%%%%%%%%%%%%%%%%%%%%%%%%%%%%%%%%%%%%%%%%%%%%%%%%%
\begin{table}
\caption {One-sided power spectral densities, $ S_N$, for the interferometers
considered in this paper.  For each detector $ S_N$ is given as a function
of the dimensionless frequency $x=\nu/\nu_0$, and is considered to be
infinite below the seismic cutoff frequency $\nu_s$.
}
\label{table2}
\bigskip
\begin {tabular}{c|c|c|c}
Detector  	&$\nu_s/$Hz &$\nu_0/$Hz &$10^{46}\times S_N(x)/$Hz$^{-1}$ \\
\tablerule
VIRGO I	&20	&500	
&$3.24\left[\left(6.23x\right)^{-5}+2x^{-1}+1+x^2\right]$ \\
LIGO I	&40	&150	
&$9\left[\left(4.49x\right)^{-56}+0.16x^{-4.52}+0.52+0.32x^2\right]$ \\
EURO	&10	&1000	
&$10^{-4}\left[0.0036x^{-4}+0.13x^{-2}+1.3\left(1+x^2\right)\right]$ \\
EURO-X	&10	&1000	&$10^{-4}\left[0.0036x^{-4}+0.13x^{-2}\right]$\\
\end {tabular}
\end {table}
%%%%%%%%%%%%%%%%%%%%%%%%%%%%%%%%%%%%%%%%%%%%%%%%%%%%%%%%%%%%%%%%%%%%%%%%%%%
%%%%%%%%%%%%%%%%%%%%%%%%%%%%%%%%%%%%%%%%%%%%%%%%%%%%%%%%%%%%%%%%%%%
\begin {table}
\caption{Ambiguity function maximized over all parameters
({\it Effectualness }) using as templates the Pad\'e-6 approximant ($P_6$) 
and the  black hole signal (BH), assuming that the two inspiralling masses are equal.
The  {\it true } signal is that emitted by the five models of neutron 
stars given in Table \ref{table1}  [6].
Values quoted are the  minimax  overlaps.
}
\label{table3}
\bigskip
\begin {tabular}{c|cc|cc|cc|cc}
     & \multicolumn{2}{c}{LIGO I} & \multicolumn{2}{c}{VIRGO I} 
     & \multicolumn{2}{c}{EURO}  
     & \multicolumn{2}{c}{EURO-X}\\
\tablerule
     &   $\left<P_6,X\right> $
     &   $\left<{BH},X\right> $
     &   $\left<P_6,X\right> $
     &   $\left<{BH},X\right> $
     &   $\left<P_6,X\right> $
     &   $\left<{BH},X\right> $ 
     &   $\left<P_6,X\right> $
     &   $\left<{BH},X\right> $\\
\tablerule
 A & 0.971 & 0.972 & 0.911  &  0.913 &  0.857  &  0.860 &  0.601 &  0.603 \\[\up]
 B & 0.984 & 0.984 & 0.965  &  0.968 &  0.933  &  0.935 &  0.755 &  0.756 \\[\up]
 C & 0.992 & 0.993 & 0.982  &  0.984 &  0.968  &  0.970 &  0.871 &  0.873 \\[\up]
 D & 0.999 & 0.999 & 0.997  &  0.998 &  0.997  &  0.998 &  0.998 &  0.998 \\[\up]
 E & 0.999 & 1.000 & 0.995  &  0.999 &  0.996  &  0.999 &  0.998 &  0.999 \\[\up]
\end {tabular}
\end {table}

%%%%%%%%%%%%%%%%%%%%%%%%%%%%%%%%%%%%%%%%%%%%%%%%%%%%%%%%%%%%%%%%%%%
%%%%%%%%%%%%%%%%%%%%%%%%%%%%%%%%%%%%%%%%%%%%%%%%%%%%%%%%%%%%%%%%%%%%%%%%%%%
\begin {table}
\caption{Ambiguity function maximized over  the phases and the time lag 
({\it Faithfulness}). Templates and true signals are chosen  as in Table II.
Values quoted are the {\it minimax} overlaps.} 
\label{table4}
\bigskip
\begin{tabular}{c|cc|cc|cc|cc}
     & \multicolumn{2}{c}{LIGO I} & \multicolumn{2}{c}{VIRGO I} 
     & \multicolumn{2}{c}{EURO} 
     & \multicolumn{2}{c}{EURO-Xylo}\\      
\tablerule
     &   $\left <P_6,X \right > $ 
     &   $\left <{BH},X \right > $ 
     &   $\left <P_6,X \right > $ 
     &   $\left <{BH},X \right > $ 
     &   $\left <P_6,X \right > $ 
     &   $\left <{BH},X \right > $
     &   $\left <P_6,X \right > $
     &   $\left <{BH},X \right > $\\
\tablerule
 A &  0.955  &  0.957  &  0.867  &  0.869 & 0.687  &  0.705 &  0.538 &  0.546   \\[\up] 
 B &  0.977  &  0.980  &  0.923  &  0.924 & 0.769  &  0.766 &  0.593 &  0.596   \\[\up] 
 C &  0.989  &  0.991  &  0.945  &  0.953 & 0.883  &  0.894 &  0.792 &  0.801   \\[\up] 
 D &  0.998  &  0.997  &  0.994  &  0.986 & 0.986  &  0.984 &  0.923 &  0.979   \\[\up] 
 E &  0.998  &  1.000  &  0.992  &  0.999 & 0.974  &  0.999 &  0.969 &  0.999   \\[\up] 
\end{tabular}
\end{table}
%%%%%%%%%%%%%%%%%%%%%%%%%%%%%%%%%%%%%%%%%%%%%%%%%%%%%%%%%%%%%%%%%%%%%%%%%%%

%%%%%%%%%%%%%%%%%%%%%%%%%%%%%%%%%%%%%%%%%%%%%%%%%%%%%%%%%%%%%%%%%%%%%%%%%%%
%%%%%%%%%%%%%%%%%%%%%%%%%%%%%%%%%%%%%%%%%%%%%%%%%%%%%%%%%%%%%%%%%%%%%%%%%%%
%%%%%%%%%%%%%%%%%%%%%%%%%   FIGURES   %%%%%%%%%%%%%%%%%%%%%%%%%%%%%%%%%%%%%
%%%%%%%%%%%%%%%%%%%%%%%%%%%%%%%%%%%%%%%%%%%%%%%%%%%%%%%%%%%%%%%%%%%%%%%%%%%
%%%%%%%%%%%%%%%%%%%%%%%%%%%%%%%%%%%%%%%%%%%%%%%%%%%%%%%%%%%%%%%%%%%%%%%%%%%
%%%%%%%%%%%%%%%%%%%%%%%%%%%%%%%%%%%%%%%%%%%%%%%%%%%%%%%%%%%%%%%%%%%%%%%

%%%%%%%%%%%%%%%%%%%%%%%%%%%%%%%%%%%%%%%%%%%%%%%%%%%%%%%%%%%%%%%%%%%%%
%%%%%%%%%%%%%%%%%%%%%%%%%%%%%%%%%%%%%%%%%%%%%%%%%%%%%%%%%%%%%%%%%%%%%

%%%%%%%%%%%%%%%%%%%%%%%%%%%%%%%%%%%%%%%%%%%%%%%%%%%%%%%%%%%%%%%%%%%%%
%%%%%%%%%%%%%%%%%%%%%%%%%%%%%%%%%%%%%%%%%%%%%%%%%%%%%%%%%%%%%%%%%%%%%

%%%%%%%%%%%%%%%%%%%%%%%%%%%%%%%%%%%%%%%%%%%%%%%%%%%%%%%%%%%%%%%%%%%%%%%
% References
%%%%%%%%%%%%%%%%%%%%%%%%%%%%%%%%%%%%%%%%%%%%%%%%%%%%%%%%%%%%%%%%%%%%%%%

\end{document}